\documentclass[twocolumn,10pt]{asme2e}

\usepackage{amsmath, amssymb}

\newtheorem{rem}{Remark}

\usepackage{graphicx}
\usepackage{subfigure}
\usepackage{psfrag}

\confshortname{OMAE2008}
\conffullname{27th International Conference on Offshore Mechanics and Arctic Engineering}
\confdate{15-20}
\confmonth{June}
\confyear{2008}
\confcity{Estoril}
\confcountry{Portugal}

\papernum{OMAE2008-57060}

\title{Simulation of Free Surface Compressible Flows Via a Two Fluid Model}

\author{Fr\'{e}d\'{e}ric Dias\thanks{Address all correspondence to this author.}, Denys Dutykh, Jean-Michel Ghidaglia
    \affiliation{
    	\'{E}cole Normale Sup\'{e}rieure de Cachan\\
    	Centre de Math\'{e}matiques et de Leurs Applications\\
    	61, Avenue du Pr\'{e}sident Wilson\\
    	94235 Cachan France\\
    	Email: dias@cmla.ens-cachan.fr, Denys.Dutykh@cmla.ens-cachan.fr, jmg@cmla.ens-cachan.fr}
}

\newcommand{\od}[2]{\frac{d#1}{d#2}}
\newcommand{\pd}[2]{\frac{\partial#1}{\partial#2}}
\newcommand{\set}[1]{\left\{ #1 \right\}}
\newcommand{\vol}{\mathop{\mathrm{vol}}}
\newcommand{\area}{\mathop{\mathrm{area}}}
\newcommand{\sign}{\mathop{\mathrm{sign}}}
\newcommand{\diag}{\mathop{\mathrm{diag}}}

\def\g{\vec{g}}
\def\u{\vec{u}}

\def\div{\nabla\cdot}
\def\I{\mathbb{I}}
\def\w{\mathbf{w}}
\def\F{\mathcal{F}}
\def\S{\mathcal{S}}

\def\n{\vec{n}}
\def\A{\mathbb{A}}
\def\N{\mathcal{N}}

\def\x{\vec{x}}

\def\R{\mathbb{R}}

\def\T{\mathcal{T}}
\def\O{\mathcal{O}}
\def\grad{\nabla}
\def\L{\mathcal{L}}

\begin{document}

\maketitle

\begin{abstract}
{\it The purpose of this communication is to discuss the simulation of a free surface compressible flow between two fluids, typically air and water. We use a two fluid model with the same velocity, pressure and temperature for both phases. In such a numerical model, the free surface becomes a thin three dimensional zone. The present method has at least three advantages: (i) the free-surface treatment is completely implicit; (ii) it can naturally handle wave breaking and other topological changes in the flow; (iii) one can easily vary the Equation of States (EOS) of each fluid (in principle, one can even consider tabulated EOS). Moreover, our model is unconditionally hyperbolic for reasonable EOS.}
\end{abstract}

\section*{Introduction}

One of the challenges in Computational Fluid Dynamics (CFD) is to determine efforts exerted by waves on coastal structures. Such flows can be quite complicated and in particular when the sea is rough, wave breaking can lead to flows that cannot be described by basic models like {\it e.g.} free surface Euler or Navier-Stokes equations. In a free surface model, the boundary between the gas (air) and the liquid (water) is a surface. The liquid flow is assumed to be incompressible, while the gas is represented by a media, above the liquid, where the pressure is constant (the atmospheric pressure in general). Such a description is known to be valid as far as propagation in the open sea of waves with moderate amplitude is concerned. Clearly it is not satisfactory when waves either break or hit coastal structures like offshore platforms, jetties, piers, breakwaters {\it etc.} \ldots

In this work, our goal is to investigate a two-fluid model for this kind of problem. It belongs to the family of averaged models, that is although the two fluids considered are not miscible, there exists a length scale $\epsilon$ in order that each averaging volume (of size $\epsilon^3$) contain representative samples of each of the fluids. Once the averaging process is done, it is assumed that the two fluids share, locally, the same pressure, temperature and velocity. Such models are called homogeneous models in the literature. They can be seen as limiting case of more general two-fluid models where the two fluids could have different temperatures and velocities \cite{Ishii1975}.

The influence of the presence of air in wave impacts is a difficult topic. While it is usually thought that the presence of air softens the impact pressures, recent results by Bullock et al. \cite{Bullock2007} show that the cushioning effect due to aeration via the increased compressibility of the air-water mixture is not necessarily a dominant effect. First of all, air may become trapped or entrained in the water in different ways, for example as a single bubble trapped against a wall, or as a column or cloud of small bubbles. In addition, it is not clear which quantity is the most appropriate to measure impacts. For example some researchers pay more attention to the pressure impulse than to pressure peaks. The pressure impulse is defined as the integral of pressure over the short duration of impact. Bagnold \cite{Bagnold1939}, for example, noticed  that the maximum pressure and impact duration differed from one identical wave impact to the next, even in carefully controlled laboratory experiments, while the pressure impulse appears to be more repeatable. For sure, the simple one-fluid models which are commonly used for examining the peak impacts are no longer appropriate in the presence of air. There are few studies dealing with two-fluid models. An exception is the work by Peregine and his collaborators. Wood, Peregrine \& Bruce \cite{Wood2000} used the pressure impulse approach to model a trapped air pocket. Peregrine \& Thais \cite{Peregrine1996} examined the effect of entrained air on a particular kind of violent water wave impact by considering a filling flow. Bullock et al. \cite{Bullock2001} found pressure reductions when comparing wave impact between fresh and salt water where, due to the different properties of the bubbles in the two fluids, the aeration levels are much higher in salt water than in fresh. H. Bredmose recently performed numerical experiments on a two-fluid system which is quite similar to the one we will use below. He developed a finite volume solver for aerated flows named Flair \cite{Peregrine2004}.

\section*{Mathematical model}

In this section we present the equations which govern the motion of two phase mixtures in a computational domain $\Omega$. First of all, we need to introduce the notation which will be used throughout this article. We use superscripts $\pm$ to denote any quantity which is related to liquid and gas respectively. For example, $\alpha^+$ and $\alpha^-$ denote the volume fraction of liquid and gas and obviously satisfy the condition $\alpha^+ + \alpha^-=1$. Then, we have the following classical quantities: $\rho^\pm$, $\u$, $p$, $e$, $E$, $\g$ which denote the density of each phase, the velocity field vector, the pressure, the internal \& total energy and the acceleration due to gravity correspondingly.

Conservation of mass (one equation for each phase), momentum and energy lead to the four following equations:
\begin{eqnarray}\label{eq:massphys}
  (\alpha^\pm\rho^\pm)_t  + \div(\alpha^\pm\rho^\pm\u) &=& 0, \\
  (\rho\u)_t + \div\bigl(\rho\u\otimes\u + p\I\bigr) &=& \rho\g, \\
  \bigl(\rho E\bigr)_t + \div\bigl(\rho H\u\bigr) &=& \rho\g\cdot\u, \label{eq:energyphys}
\end{eqnarray}
where $\rho := \alpha^+\rho^+ + \alpha^-\rho^-$ (the total density), $H := E + \frac{p}{\rho}$ (the specific enthalpy), $E = e + \frac12|\u|^2$ (the total energy). This system can be seen as the single energy and infinite drag limit of the more conventional six equations model \cite{Ishii1975}. The above system contains five unknowns $\alpha^\pm\rho^\pm$, $\u$, $p$ and $E$ and only four governing equations (\ref{eq:massphys}) - (\ref{eq:energyphys}). In order to close the system, we need to provide the so-called equation of state (EOS) $p=p^\pm(\rho^\pm,e^\pm)$. The construction of the EOS will be discussed below.

It is possible to rewrite these equations as a system of balance laws
\begin{equation}\label{eq:conslaws}
  \pd{\w}{t} + \div\F(\w) = S(\w),
\end{equation}
where $\w(x,t):\R^d\times\R^+\mapsto \R^m$ is the vector of conservative variables (in the present study $d=2$ or $3$ and $m=5$), $\F(\w)$ is the advective flux function and $S(\w)$ the source term.

The conservative variables in the 2D case are defined as follows:
\begin{equation}\label{eq:consvars}
  \w = (w_i)_{i=1}^{5} := (\alpha^+\rho^+, \alpha^-\rho^-,\;\;\rho u,\;\;\rho v,\;\;\rho E).
\end{equation}
The flux projection on the normal direction $\n=(n_1, n_2)$ can be expressed in physical and conservative variables
\begin{multline}\label{eq:normalAdvflux}
  \F\cdot\n = (\alpha^+\rho^+u_n, \alpha^-\rho^-u_n, \rho u u_n + p n_1, \rho v u_n + pn_2, \rho H u_n) =\\ \Bigl(w_1\frac{w_3n_1 + w_4n_2}{w_1+w_2}, w_2\frac{w_3n_1 + w_4n_2}{w_1+w_2}, w_3\frac{w_3n_1 + w_4n_2}{w_1+w_2} + p n_1, \\ w_4\frac{w_3n_1 + w_4n_2}{w_1+w_2} + pn_2, (w_5 + p)\frac{w_3n_1 + w_4n_2}{w_1+w_2}\Bigr)
\end{multline}
where $u_n := \u\cdot\n = un_1 + vn_2$ is the velocity projection on the normal direction $\n$. The jacobian matrix $\A_n(\w) := \pd{(\F\cdot\n)(\w)}{\w}$ can be easily computed. 
This matrix has three distinct eigenvalues:
\begin{equation*}
  \lambda_1 = u_n - c_s, \quad
  \lambda_{2,3,4} = u_n, \quad
  \lambda_5 = u_n + c_s,
\end{equation*}
where $c_s$ is the sound speed in the mixture. Its expression can be found in \cite{Dias2008}. One can conclude that the system (\ref{eq:massphys}) -- (\ref{eq:energyphys}) is hyperbolic. This hyperbolicity represents the major advantage of this model. The computation of the eigenvectors is trickier but can still be performed analytically. We do not give here the final expressions since they are cumbersome.

\subsection*{Equation of state}

In the present work we assume that the light fluid is described by an ideal gas type law
\begin{equation}\label{eq:light}
  p^- = (\gamma - 1) \rho^- e^-, \qquad e^- = c_v^- T^-,
\end{equation}
while the heavy fluid is modeled by Tait's law. In the literature Tait's law is sometimes called the stiffened gas law \cite{Godunov1979, Harlow1971}:
\begin{equation}\label{eq:heavy}
  p^+ + \pi_0 = (\N - 1) \rho^+ e^+, \qquad e^+ = c_v^+T^+ + \frac{\pi_0}{\N \rho^+},
\end{equation}
where the quantities $\gamma$, $c_v^\pm$, $\N$, $\pi_0$ are constants. For example, pure water is well described when we take $\N = 7$ and $\pi_0 = 2.1\times10^9$ Pa.

\begin{rem}
  In practice, the constants $c_v^\pm$ can be calculated after simple algebraic manipulations of equations (\ref{eq:light}), (\ref{eq:heavy}) and matching with experimental values at normal conditions:
  \begin{equation*}
    c_v^- \equiv \frac{p_0}{(\gamma-1)\rho^-_0 T_0}, \quad
    c_v^+ \equiv \frac{\N p_0 + \pi_0}{(\N-1)\N\rho_0^+ T_0}.
  \end{equation*}
\end{rem}


The sound velocities in each phase are given by the following formulas:
\begin{equation}\label{eq:soundspeed}
  (c_s^-)^2 = \frac{\gamma p^-}{\rho^-}, \qquad
  (c_s^+)^2 = \frac{\N p^+ + \pi_0}{\rho^+}.
\end{equation}

In order to construct an equation of state for the mixture, we make the additional assumption that the two phases are in thermodynamic equilibrium:
\begin{equation}
  p^+ = p^-, \qquad T^+ = T^-.
\end{equation}
Below, values of the common pressure and common temperature will be denoted by $p$ and $T$ respectively. The technical details can be found in Chapter 3, \cite{Dutykh2007a}.

\section*{Finite volume scheme on unstructured meshes}

Finite volume methods are a class of discretization schemes that have proven highly successful in solving numerically a wide class of conservation law systems. These systems often come from compressible fluid dynamics. When compared to other discretization methods such as finite elements or finite differences, the primary interests of finite volume methods are robustness, applicability on very general unstructured meshes, and the intrinsic local conservation properties. Hence, with this type of discretization, we conserve ``exactly'' the mass, momentum and total energy\footnote{This statement is true in the absence of source terms and appropriate boundary conditions.}.

In order to solve numerically the system of balance laws (\ref{eq:massphys}) -- (\ref{eq:energyphys}) we use (\ref{eq:conslaws}). The system (\ref{eq:conslaws}) should be provided with initial condition
\begin{equation}\label{eq:initialcond}
  \w(x,0) = \w_0(x)
\end{equation}
and appropriate boundary conditions.

\begin{figure}[htbp]
\centering
\psfrag{O}{$O$}
\psfrag{K}{$K$}
\psfrag{M}{$\partial K$}
\psfrag{n}{$\n_{KL}$}
\psfrag{L}{$L$}
\includegraphics[width=5cm]{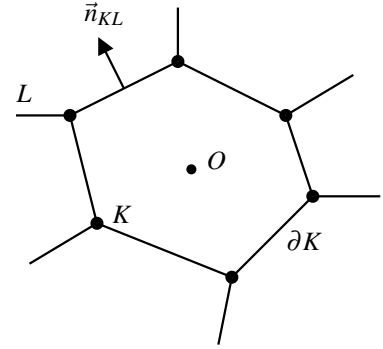}
\caption[An example of control volume $K$.]{An example of control volume $K$ with barycenter $O$. The normal pointing from $K$ to $L$ is denoted by $\n_{KL}$.}
\label{fig:controlvol}
\end{figure}

The computational domain $\Omega\subset\R^d$ is triangulated into a set of non overlapping control volumes that completely cover the domain. Let $\T$ denote a tesselation of the domain $\Omega$ with control volume $K$ such that
\begin{equation*}
  \cup_{K\in\T} \bar{K} = \bar{\Omega}, \quad \bar{K} := K \cup \partial K.
\end{equation*}
For two distinct control volumes $K$ and $L$ in $\T$, the intersection is either an edge (2D) or face (3D) with oriented normal $\n_{KL}$ or else a set of measure at most $d-2$ (in 2D it is just a vertex, in 3D it can also be a segment, for example). We need to introduce the following notation for the neighbourhood of $K$:
\begin{equation*}
  \N(K) := \set{L\in\T: \area(K\cap L) \neq 0},
\end{equation*}
a set of all control volumes $L$ which share a face (or an edge in 2D) with the given volume $K$. In this article, we denote by $\vol(\cdot)$ and $\area(\cdot)$ the $d$ and $d-1$ dimensional measures\footnote{In other words, in 3D the notation $\area(\cdot)$ and $\vol(\cdot)$ are very natural and mean area and volume respectively, while in 2D they refer to the area and the length.} respectively.

The choice of control volume tesselation is flexible in the finite volume method. In the present study we retained a so-called cell-centered approach, which means that degrees of freedom are associated to cell barycenters.

The first steps in Finite Volume (FV) methods are classical. We start by integrating equation (\ref{eq:conslaws}) on the control volume $K$ (see \figurename~\ref{fig:controlvol} for illustration) and we apply Gauss-Ostrogradsky theorem for advective fluxes. Then, in each control volume, an integral conservation law statement is imposed:
\begin{equation}\label{eq:conservlaw}
	\od{}{t}\int_{K} \w \;d\Omega + \int_{\partial K}\F(\w)\cdot\n_{KL} \;d\sigma 
  	  = \int_{K}S(\w) \;d\Omega\;.
\end{equation}
Physically an integral conservation law asserts that the rate of change of the total amount of a quantity (for example: mass, momentum, total energy, etc) with density $\w$ in a fixed control volume $K$ is balanced by the flux $\F$ of the quantity through the boundary $\partial K$ and the production of this quantity $\S$ inside the control volume.

The next step consists in introducing the so-called control volume cell average for each $K\in\T$
\begin{equation*}
 \w_K(t) := \frac{1}{\vol(K)}\int_{K} \w(\x,t) \;d\Omega \;.
\end{equation*}
After the averaging step, the finite volume method can be interpreted as producing a system of evolution equations for cell averages, since 
\begin{equation*}
  \od{}{t}\int_{K} \w(\x,t) \;d\Omega = \vol(K)\od{\w_K}{t} \;.
\end{equation*}
Godunov was the first \cite{Godunov1959} who pursued and applied these ideas to the discretization of the gas dynamics equations.

However, the averaging process implies piecewise constant solution representation in each control volume with value equal to the cell average. The use of such representation makes the numerical solution multivalued at control volume interfaces. Thereby the calculation of the fluxes $\int_{\partial K}\F(\w)\cdot\n_{KL} \;d\sigma$ at these interfaces is ambiguous. The next fundamental aspect of finite volume methods is the idea of substituting the true flux at interfaces by a numerical flux function
\begin{equation*}
  \left.\bigl(\F(\w)\cdot\n\bigr)\right|_{\partial K\cap\partial L} \longleftarrow \Phi(\w_K, \w_L; \n_{KL}) : \R^m\times\R^m \mapsto \R^m \;,
\end{equation*}
a Lipschitz continuous function of the two interface states $\w_K$ and $\w_L$. The heart of the matter in finite volume method consists in the choice of the numerical flux function $\Phi$. In general this function is calculated as an exact or even better approximate local solution of the Riemann problem posed at these interfaces. In the present study we decided to choose the numerical flux function according to FVCF scheme extensively described in \cite{Ghidaglia2001}.

The numerical flux is assumed to satisfy the properties:
\begin{description}
  \item[Conservation.] This property ensures that fluxes from adjacent control volumes sharing an interface exactly cancel when summed. This is achieved if the numerical flux function satisfies the identity
  \begin{equation*}
    \Phi(\w_K,\w_L; \n_{KL}) = - \Phi(\w_L,\w_K; \n_{LK}).
  \end{equation*}
  \item[Consistency.] The consistency is obtained when the numerical flux with identical state arguments (in other words it means that the solution is continuous through an interface) reduces to the true flux of the same state, i.e.
  \begin{equation*}
    \Phi(\w,\w; \n) = (\F\cdot\n)(\w).
  \end{equation*}
\end{description}

After introducing the cell averages $\w_K$ and numerical fluxes into (\ref{eq:conservlaw}), the integral conservation law statement becomes
\begin{multline*}
	\od{\w_K}{t} + \sum_{L\in\N(K)} \frac{\area(L\cap K)}{\vol(K)} \Phi(\w_K, \w_L; \n_{KL}) = \\
  	  = \frac{1}{\vol(K)} \int_{K}S(\w) \;d\Omega\;.
\end{multline*}
We denote by $S_K$ the approximation of the following quantity $\frac{1}{\vol(K)} \int_{K} S(\w) \;d\Omega$. Thus, the following system of ordinary differential equations (ODE) is called a semi-discrete finite volume method:
\begin{equation}\label{eq:semidiscrete1}
	\od{\w_K}{t} + \sum_{L\in\N(K)} \frac{\area(L\cap K)}{\vol(K)} \Phi(\w_K, \w_L; \n_{KL}) =  S_K, \quad \forall K\in\T\;.
\end{equation}
The initial condition for this system is given by projecting (\ref{eq:initialcond}) onto the space of piecewise constant functions
\begin{equation*}
  \w_K(0) = \frac{1}{\vol{(K)}} \int_{K} \w_0(x) \; d\Omega\;.
\end{equation*}

This system of ODE (\ref{eq:semidiscrete1}) should also be discretized. There is a variety of explicit and implicit time integration methods. We chose the following third order four-stage SSP-RK(3,4) scheme \cite{Shu1988, Gottlieb2001} with CFL $= 2$:
\begin{eqnarray*}
  u^{(1)} &=& u^{(n)} + \frac12\Delta t \L(u^{(n)}), \\
  u^{(2)} &=& u^{(1)} + \frac12\Delta t \L(u^{(1)}), \\
  u^{(3)} &=& \frac23u^{(n)} + \frac13u^{(2)} + \frac16\Delta t \L(u^{(n)}), \\
 	u^{(n+1)} &=& u^{(3)} + \frac12\Delta t \L(u^{(3)}).
\end{eqnarray*}

\subsection*{Sign matrix computation}

In the context of the FVCF scheme (see \cite{Ghidaglia2001} for more details), we need to compute the so-called sign matrix which is defined in the following way
\begin{equation*}
  U_n := \sign(\A_n) = R \sign(\Lambda) L,
\end{equation*}
where $R$, $L$ are matrices composed of right and left eigenvectors correspondingly, and $\Lambda = \diag(\lambda_1, \ldots, \lambda_m)$ is the diagonal matrix of eigenvalues of the Jacobian.

This definition gives the first ``direct'' method of sign matrix computation. Since the advection operator is relatively simple, after a few tricks, we can succeed in computing analytically the matrices $R$ and $L$. For more complicated two-phase models it is almost impossible to perform this computation in closed analytical form. In this case, one has to apply numerical techniques for eigensystem computations. It turns out to be costly and not very accurate. In the present work we use physical information about the model in the numerical computations.

There is another way which is less expensive. The main idea is to construct a kind of interpolation polynomial which takes the following values
\begin{equation*}
  P(u_n \pm c_s) = \sign (u_n \pm c_s), \quad
  P(u_n) = \sign (u_n).
\end{equation*}
These three conditions allow us to construct a second degree interpolation polynomial. Obviously, when $P(\lambda)$ is evaluated at $\lambda = \A_n$ we obtain the sign matrix $U_n$ as a result. The construction of the Lagrange interpolation polynomial $P(\lambda)$ is simple.

In our research code we have implemented both methods. Our experience shows that the interpolation method is quicker and gives correct results in most test cases. However, when we approach pure phase states, it shows a rather bad numerical behaviour. It can lead to instabilities and diminish overall code robustness. Thus, whenever possible we suggest to use the computation of the Jacobian eigenstructure.

\subsection*{Second order scheme}

If we analyze the above scheme, we understand that in fact, we have only one degree of freedom per data storage location. Hence, it seems that we can expect to be first order accurate at most. In the numerical community first order schemes are generally considered to be too inaccurate for most quantitative calculations. Of course, we can always make the mesh spacing extremely small but it cannot be a solution since it makes the scheme inefficient. From the theoretical point of view the situation is even worse since an $\O(h^{\frac12})$ $L_1$-norm error bound for the monotone and E-flux schemes \cite{Osher1984} is known to be sharp \cite{Peterson1991}, although an $\O(h)$ solution error is routinely observed in numerical experiments. On the other hand, Godunov has shown \cite{Godunov1959} that all linear schemes that preserve solution monotonicity are at most first order accurate. This rather negative result suggests that a higher order accurate scheme has to be essentially nonlinear in order to attain simultaneously a monotone resolution of discontinuities and high order accuracy in continuous regions.

A significant breakthrough in the generalization of finite volume methods to higher order accuracy is due to Kolgan \cite{Kolgan1972, Kolgan1975} and van Leer \cite{Leer1979}. They proposed a kind of post-treatment procedure currently known as solution \textit{reconstruction} or MUSCL\footnote{MUSCL stands for Monotone Upstream-Centered Scheme for Conservation Laws.} scheme. In the above papers the authors used linear reconstruction (it will be retained in this study as well) but this method was already extended to quadratic approximations in each cell.

In this paper we briefly describe the construction and practical implementation of a second-order nonlinear scheme on unstructured (possibly highly distorted) meshes. The main idea is to find our solution as a piecewise affine function on each cell. This kind of linear reconstruction operators on simplicial control volumes often exploit the fact that the cell average is also a pointwise value of any valid (conservative) linear reconstruction evaluated at the gravity center of a simplex. This reduces the linear reconstruction problem to that of gradient estimation at cell centers given cell averaged data. In this case, we express the reconstruction in the form
\begin{equation}\label{eq:reconstruct}
  \w_K (\x) = \bar\w_K + (\grad\w)_K\cdot(\x - \x_0), \quad K\in\T \;,
\end{equation}
where $\bar\w_K$ is the cell averaged value given by the finite volume method, $(\grad\w)_K$ is the solution gradient estimate (to be determined) on the cell $K$, $\x\in K$ and the point $\x_0$ is chosen to be the gravity center for the simplex $K$.

It is very important to note that with this type of representation (\ref{eq:reconstruct}) we remain absolutely conservative, i.e.
\begin{equation*}
  \frac{1}{\vol(K)}\int_{K} \w_K(\x)\;d\Omega \equiv \bar\w_K
\end{equation*}
due to the choice of the point $\x_0$. This point is crucial for finite volumes because of intrinsic conservative properties of this method.

In our numerical code we implemented two common techniques of gradient reconstruction: Green-Gauss integration and least squares methods. In this paper we describe only the least squares reconstruction method. The Barth-Jespersen limiter \cite{Barth1989} is incorporated to obtain non-oscillatory resolution of discontinuities and steep gradients. We refer to \cite{Dutykh2007a} for more details.

\subsection*{Least-squares gradient reconstruction method}

In this section we consider a triangle control volume $K$ with three adjacent neighbors $T_1$, $T_2$ and $T_3$. Their barycenters are denoted by $O(\x_0)$, $O_1(\x_1)$, $O_2(\x_2)$ and $O_3(\x_3)$ respectively. In the following we denote by $\w_i$ the solution value at gravity centers $O_i$:
\begin{equation*}
  \w_i := \w(\x_i), \quad \w_0 := \w(\x_0).
\end{equation*}

\begin{figure}[htbp]
\centering
\psfrag{K}{$K$}
\psfrag{T}{$T_1$}
\psfrag{S}{$T_2$}
\psfrag{U}{$T_3$}
\psfrag{O}{$O$}
\psfrag{O1}{$O_1$}
\psfrag{O2}{$O_2$}
\psfrag{O3}{$O_3$}
\includegraphics[width=6cm]{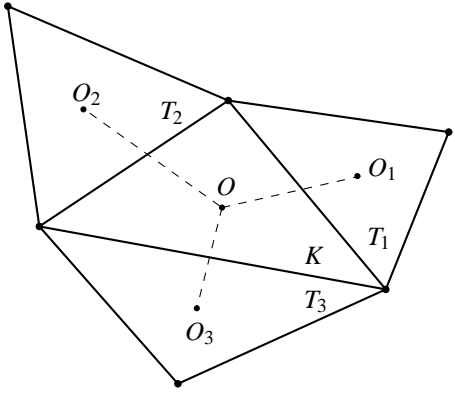}
\caption[Illustration for least-squares gradient reconstruction.]{Illustration for least-squares gradient reconstruction. We depict a triangle control volume with three adjacent neighbors.}
\label{fig:leastsq}
\end{figure}

Our purpose here is to estimate $\grad\w = (\partial_x\w, \partial_y\w)$ on the cell $K$. Using Taylor formula, we can write down the three following relations:
\begin{eqnarray}\label{eq:firstconstraint}
  \w_1 - \w_0 &=& (\grad\w)_K\cdot(\x_1 - \x_0) + \O(h^2), \\
  \w_2 - \w_0 &=& (\grad\w)_K\cdot(\x_2 - \x_0) + \O(h^2), \\
  \w_3 - \w_0 &=& (\grad\w)_K\cdot(\x_3 - \x_0) + \O(h^2). \label{eq:lastconstraint}
\end{eqnarray}
If we drop higher order terms $\O(h^2)$, these relations can be viewed as a linear system of three equations for two unknowns\footnote{This simple estimation is done for scalar case only $\w = (w)$. For more general vector problems the numbers of equations and unknowns have to be changed depending on the dimension of vector $\w$.} $(\partial_x\w, \partial_y\w)$.  This situation is due to the fact that the number of edges incident to a simplex mesh in $\R^d$ is greater or equal to $d$ thereby producing linear constraint equations (\ref{eq:firstconstraint}) -- (\ref{eq:lastconstraint}) which will be solved analytically here in a least squares sense.

First of all, each constraint (\ref{eq:firstconstraint}) -- (\ref{eq:lastconstraint}) is multiplied by a weight $\omega_i\in(0,1)$ which will be chosen below to account for distorted meshes. In matrix form our non-square system becomes
\begin{equation*}
  \begin{pmatrix}
    \omega_1\Delta x_1 & \omega_1\Delta y_1 \\
    \omega_2\Delta x_2 & \omega_2\Delta y_2 \\
    \omega_3\Delta x_3 & \omega_1\Delta y_3 \\
  \end{pmatrix}
  (\grad\w)_K = 
  \begin{pmatrix}
    \omega_1 (\w_1 - \w_0) \\
    \omega_2 (\w_2 - \w_0) \\
    \omega_3 (\w_3 - \w_0) \\
  \end{pmatrix},
\end{equation*}
where $\Delta x_i = x_i - x_0$, $\Delta y_i = y_i - y_0$. 
For further developments it is convenient to rewrite our constraints in abstract form
\begin{equation}\label{eq:abstract}
  [\vec{L_1},\; \vec{L_2}]\cdot (\grad\w)_K = \vec{f}.
\end{equation}
We use a normal equation technique in order to solve symbolically this abstract form in a least squares sense. Multiplying on the left both sides of (\ref{eq:abstract}) by $[\vec{L_1} \vec{L_2}]^t$ yields
\begin{equation}\label{eq:squaresystem}
  G(\grad\w)_K = \vec{b}, \quad 
  G = (l_{ij})_{1\leq i,j\leq 2} = 
  \begin{pmatrix}
    (\vec{L_1}\cdot\vec{L_1}) & (\vec{L_1}\cdot\vec{L_2}) \\
    (\vec{L_2}\cdot\vec{L_1}) & (\vec{L_2}\cdot\vec{L_2}) \\
  \end{pmatrix}
\end{equation}
where $G$ is the Gram matrix of vectors $\set{\vec{L_1},\vec{L_2}}$ and 
$
	\vec{b} = 
	\begin{pmatrix}
	  (\vec{L_1}\cdot\vec{f}) \\
	  (\vec{L_2}\cdot\vec{f}) \\
	\end{pmatrix}.
$
The so-called normal equation (\ref{eq:squaresystem}) is easily solved by Cramer's rule to give the following result
\begin{equation*}
  (\grad\w)_K = \frac{1}{l_{11}l_{22} - l_{12}^2}
  \begin{pmatrix}
    l_{22}(\vec{L_1}\cdot\vec{f}) - l_{12}(\vec{L_2}\cdot\vec{f}) \\
    l_{11}(\vec{L_2}\cdot\vec{f}) - l_{12}(\vec{L_1}\cdot\vec{f}) \\
  \end{pmatrix}.
\end{equation*}
The form of this solution suggests that the least squares linear reconstruction can be efficiently computed without the need for storing a non-square matrix.

Now we have to discuss the choice of weight coefficients $\set{\omega_i}_{i=1}^3$. The basic idea is to attribute bigger weights to cells barycenters closer to the node $N$ under consideration. One of the possible choices consists in taking a harmonic mean of respective distances $r_i = ||\x_i - \x_N||$. This purely metric argument takes the following mathematical form:
\begin{equation*}
  \omega_i = \frac{||\x_i - \x_N||^{-k}}{\sum_{j=1}^3||\x_j - \x_N||^{-k}},
\end{equation*}
where $k$ in practice is taken to be one or two (in our numerical code we choose $k=1$).


\section*{Numerical results: falling water column}

\begin{figure}[htbp]
\centering
\psfrag{A}{$\alpha^+ = 0.9$}
\psfrag{B}{$\alpha^- = 0.1$}
\psfrag{C}{$\alpha^+ = 0.1$}
\psfrag{D}{$\alpha^- = 0.9$}
\psfrag{0}{$0$}
\psfrag{0.3}{$0.3$}
\psfrag{0.65}{$0.65$}
\psfrag{0.7}{$0.7$}
\psfrag{0.05}{$0.05$}
\psfrag{1}{$1$}
\psfrag{0.9}{$0.9$}
\psfrag{g}{$\g$}
\includegraphics[width=9cm]{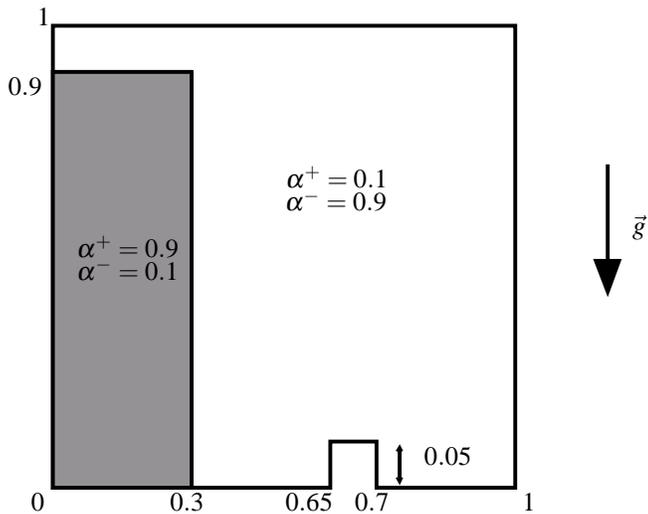}
\caption[Geometry and initial condition for falling water]{Falling water column test case. Geometry and initial condition.}
\label{fig:falling_water}
\end{figure}

\begin{figure}
	\centering
	\subfigure[$t = 0.005$]%
	{\includegraphics[width=0.2\textwidth]{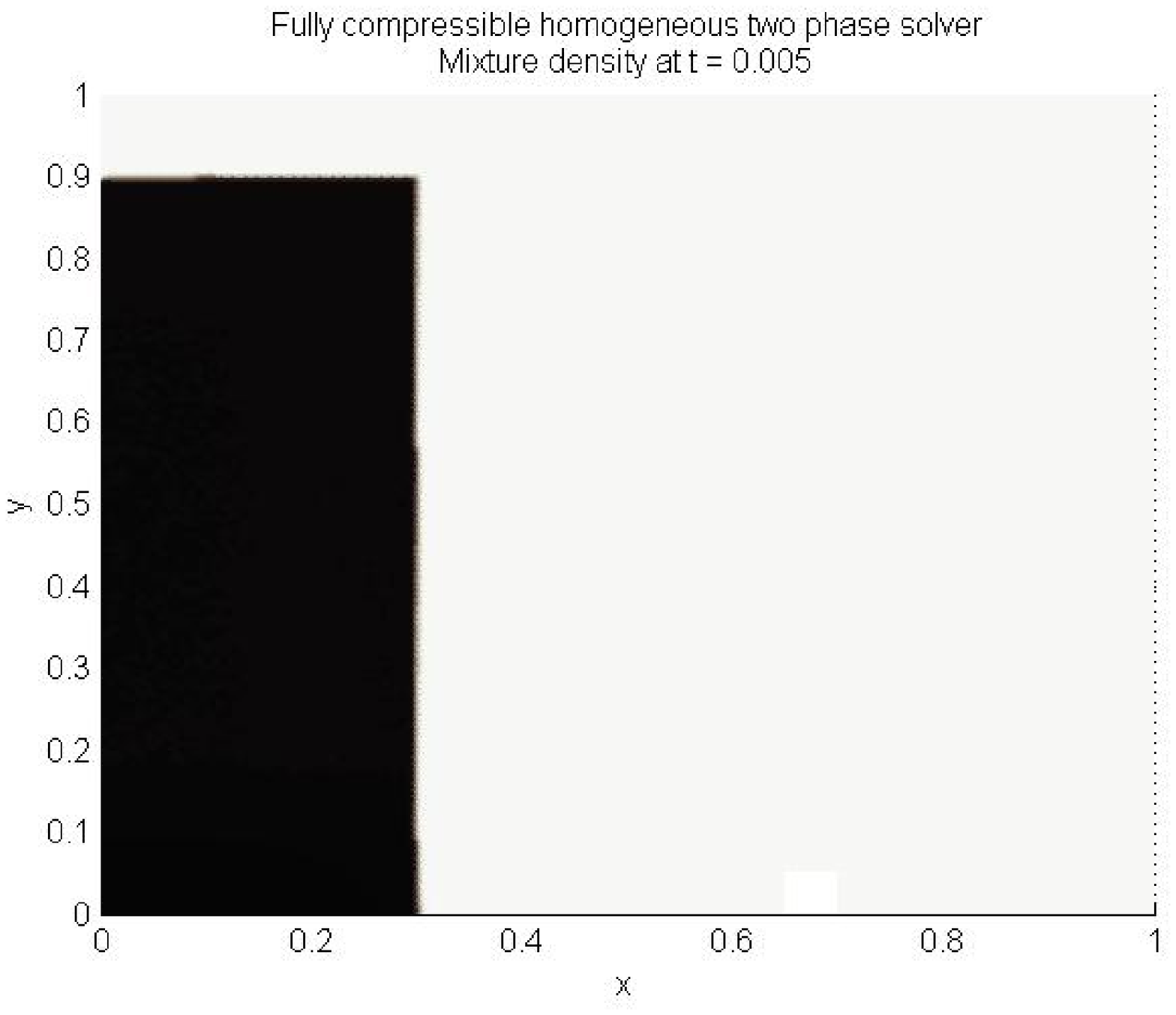}}
	\subfigure[$t = 0.06$]%
	{\includegraphics[width=0.2\textwidth]{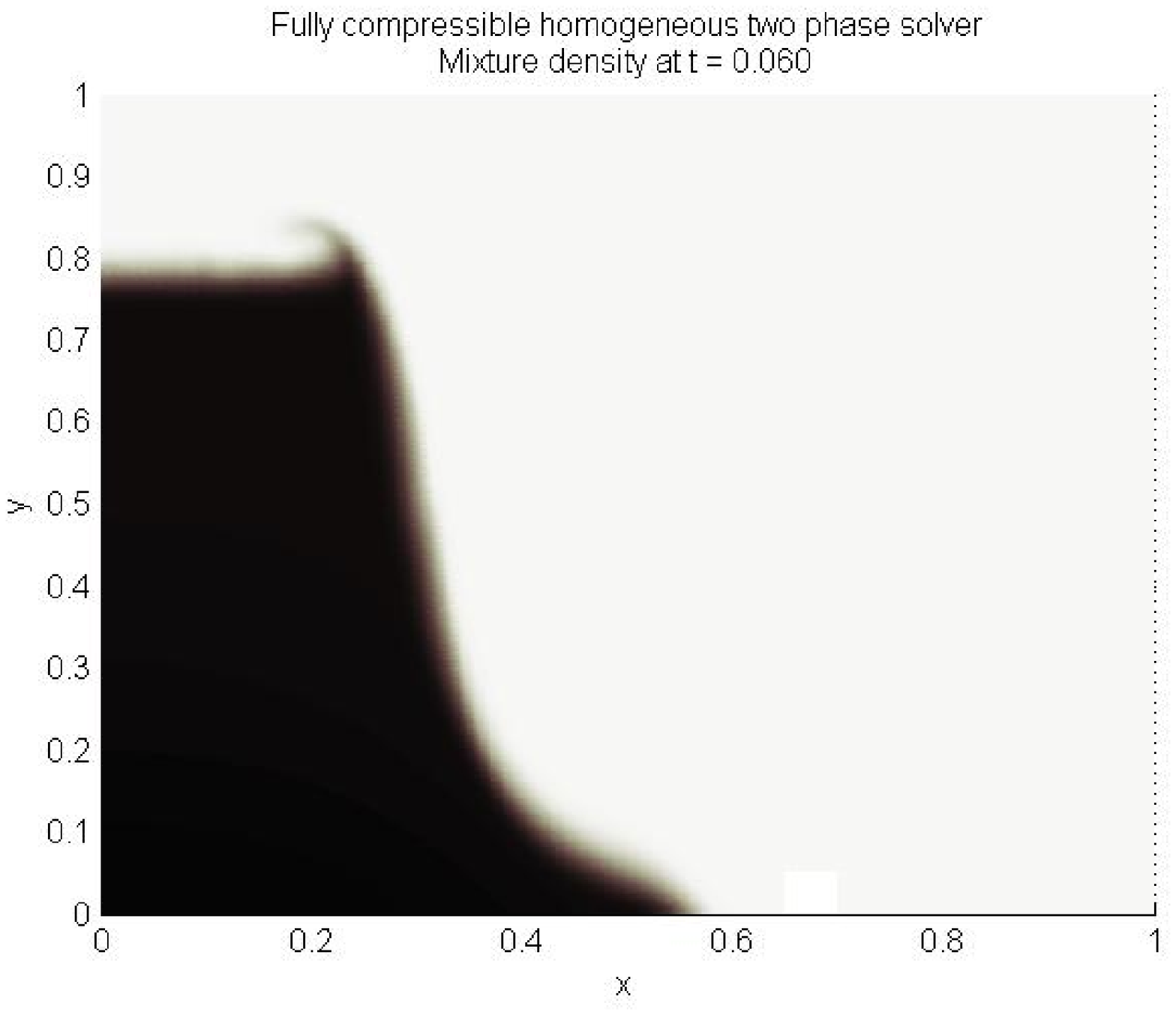}}
	\caption[The beginning of column dropping process]{Initial condition and the beginning of column dropping process.}
	\label{fig:debutSplash}
\end{figure}

\begin{figure}
	\centering
	\subfigure[$t = 0.5$]%
	{\includegraphics[width=0.2\textwidth]{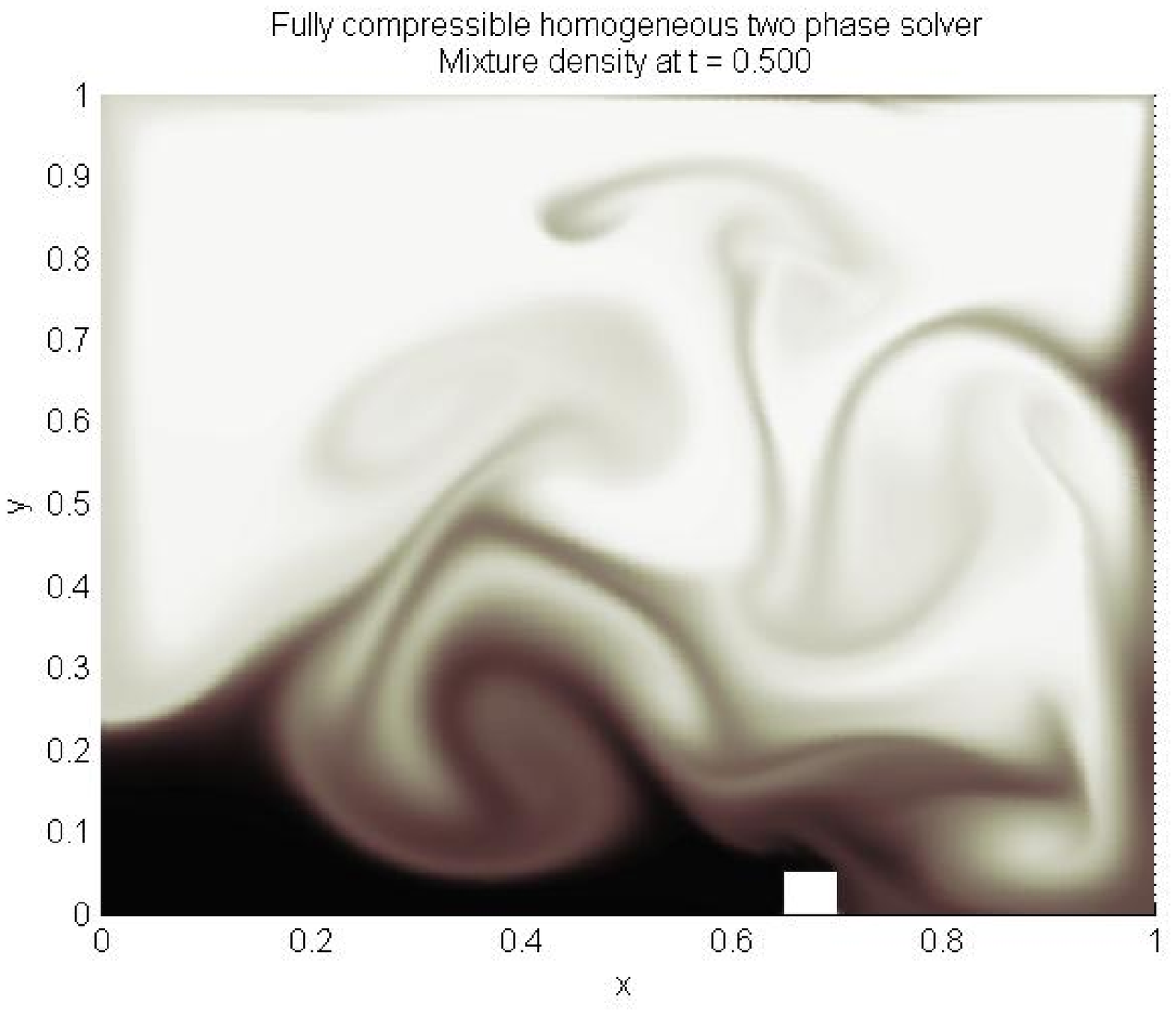}} \quad
	\subfigure[$t = 0.675$]%
	{\includegraphics[width=0.2\textwidth]{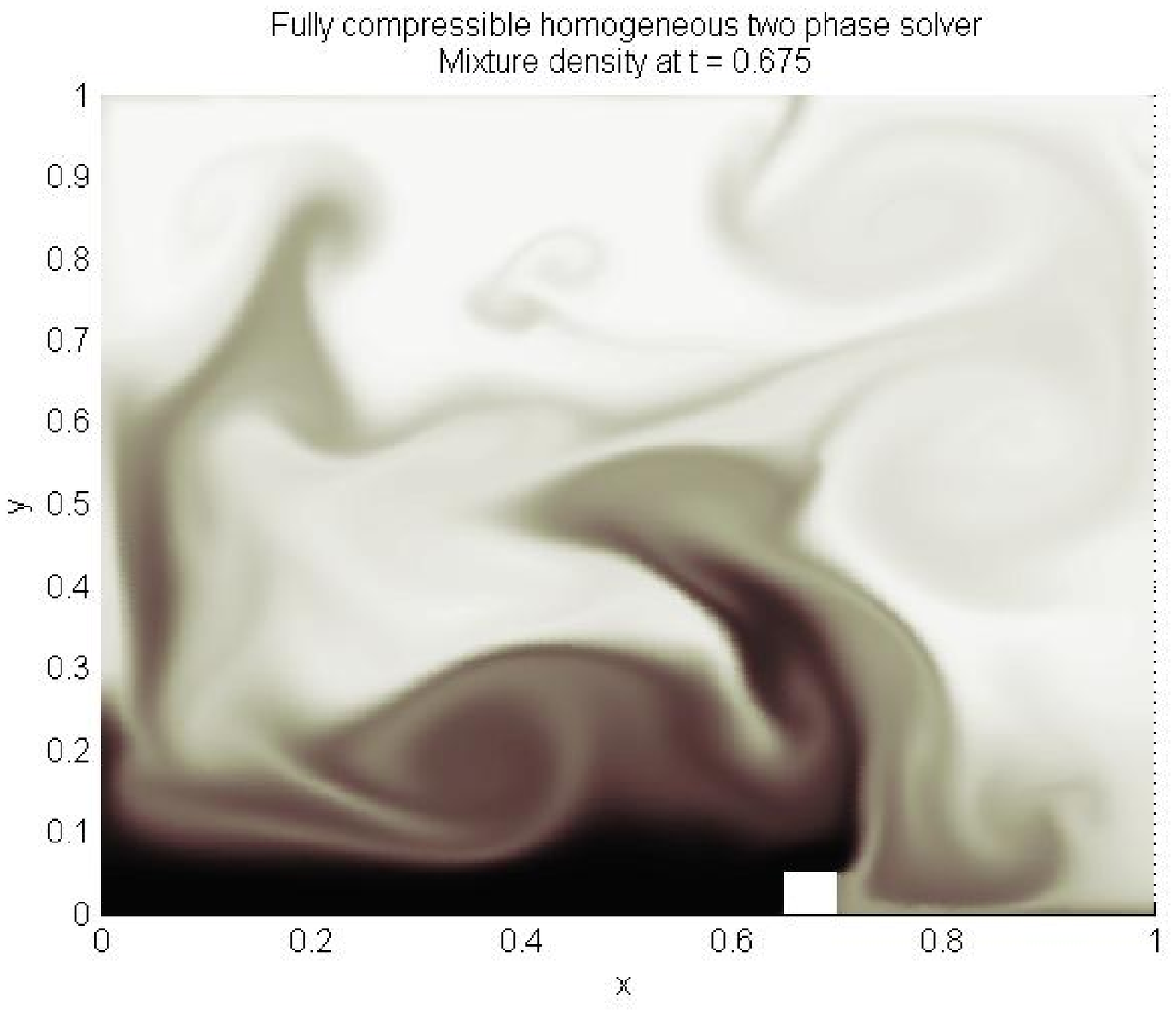}}
	\caption[Turbulent mixing process]{Turbulent mixing process.}
	\label{fig:lastSplash}
\end{figure}

\begin{figure}
	\centering
		\includegraphics[width=0.40\textwidth]{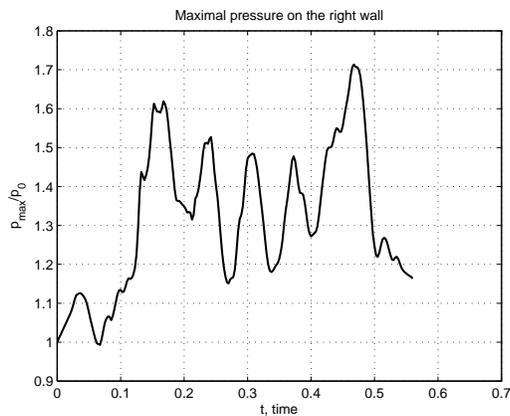}
	\caption{Maximal pressure on the right wall. Heavy gas case.}
	\label{fig:wallpress1}
\end{figure}

A classical test in violent flows is the dam break problem. This problem can be simplified as follows: a water column is released at time $t=0$ and falls under gravity. In addition, there is a step in the bottom. During its fall, the liquid hits this step and recirculation is generated behind the step. Then the liquid hits the vertical wall and climbs along this wall. The geometry and initial condition for this test case are depicted on \figurename~\ref{fig:falling_water}. Initially the velocity field is taken to be zero. The unstructured triangular grid used in this computation contained about $108 000$ control volumes (which in this case are triangles). The results of this simulation are presented on Figures \ref{fig:debutSplash} -- \ref{fig:lastSplash}. We emphasize here that there is no interface between the liquid and the gas. The dark mixture contains mostly liquid (90\%) and the light mixture contains mostly gas (90\%). An interesting quantity is the impact pressure along the wall. It is shown in \figurename~\ref{fig:wallpress1}, where the maximal pressure on the right wall is plotted as a function of time $t \mapsto \max_{(x,y) \in 1\times[0,1]} p(x,y,t)$.

\begin{figure}
	\centering
		\includegraphics[width=0.40\textwidth]{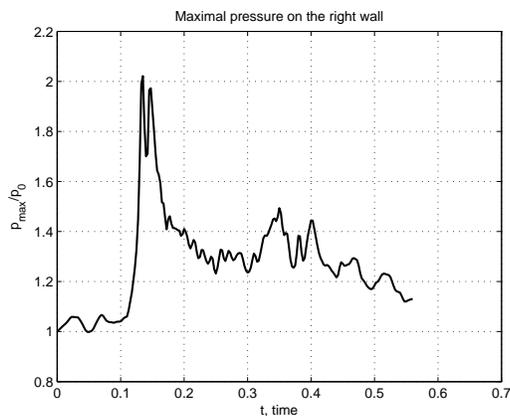}
	\caption{Maximal pressure on the right wall as a function of time. Light gas.}
	\label{fig:wallpress2}
\end{figure}

Then, we performed other computations with volume fractions closer to pure phases. For example, we show some results with the gas mixture modelled with $\alpha^+ = 0.05$, $\alpha^- = 0.95$. The pressure is recorded as well and this result is plotted on \figurename~\ref{fig:wallpress2}. One can see that the peak value is higher and the impact is more localised in time.

\begin{figure}
	\centering
	\subfigure[$t = 0.1$]%
	{\includegraphics[width=0.2\textwidth]{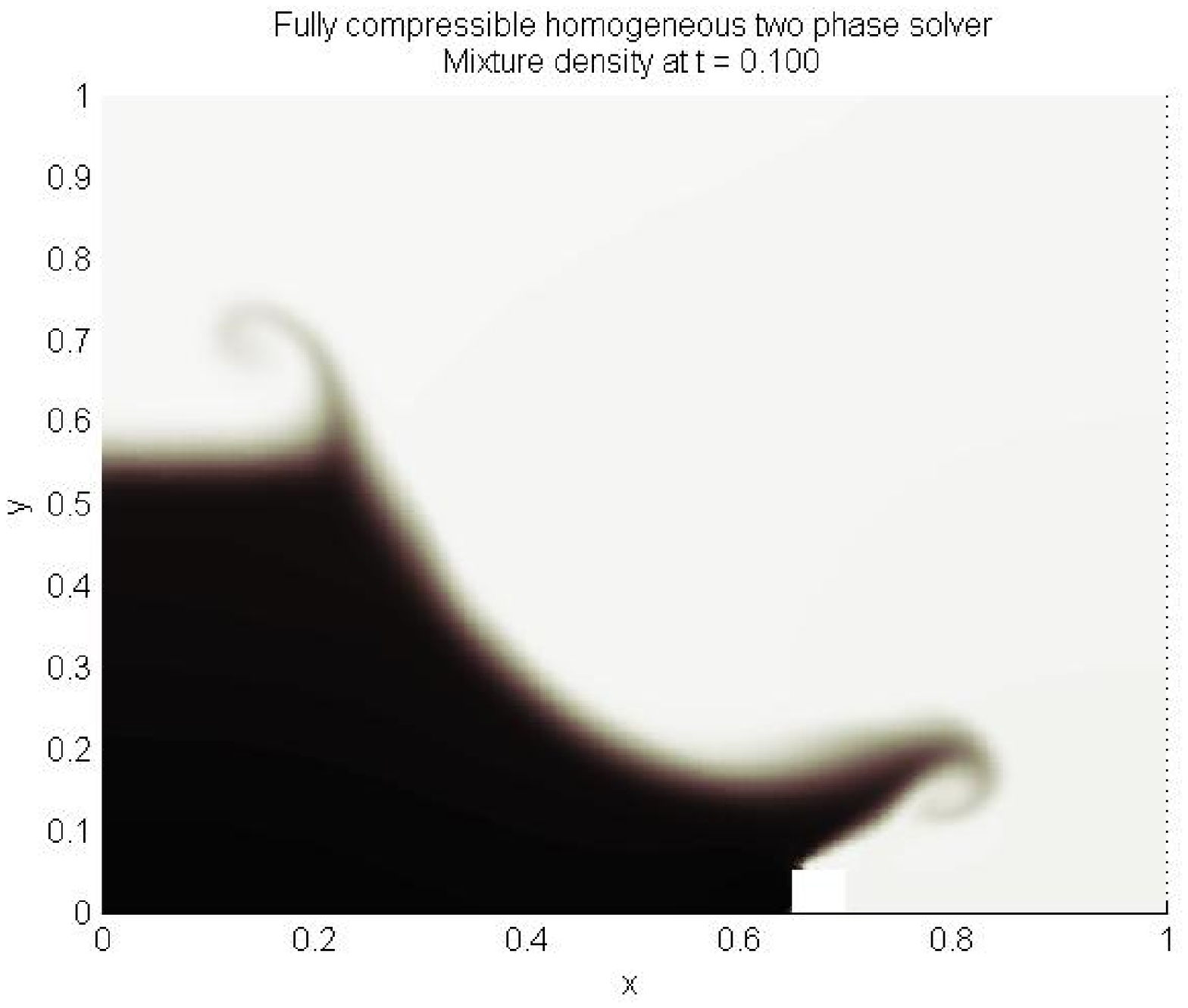}}
	\subfigure[$t = 0.125$]%
	{\includegraphics[width=0.2\textwidth]{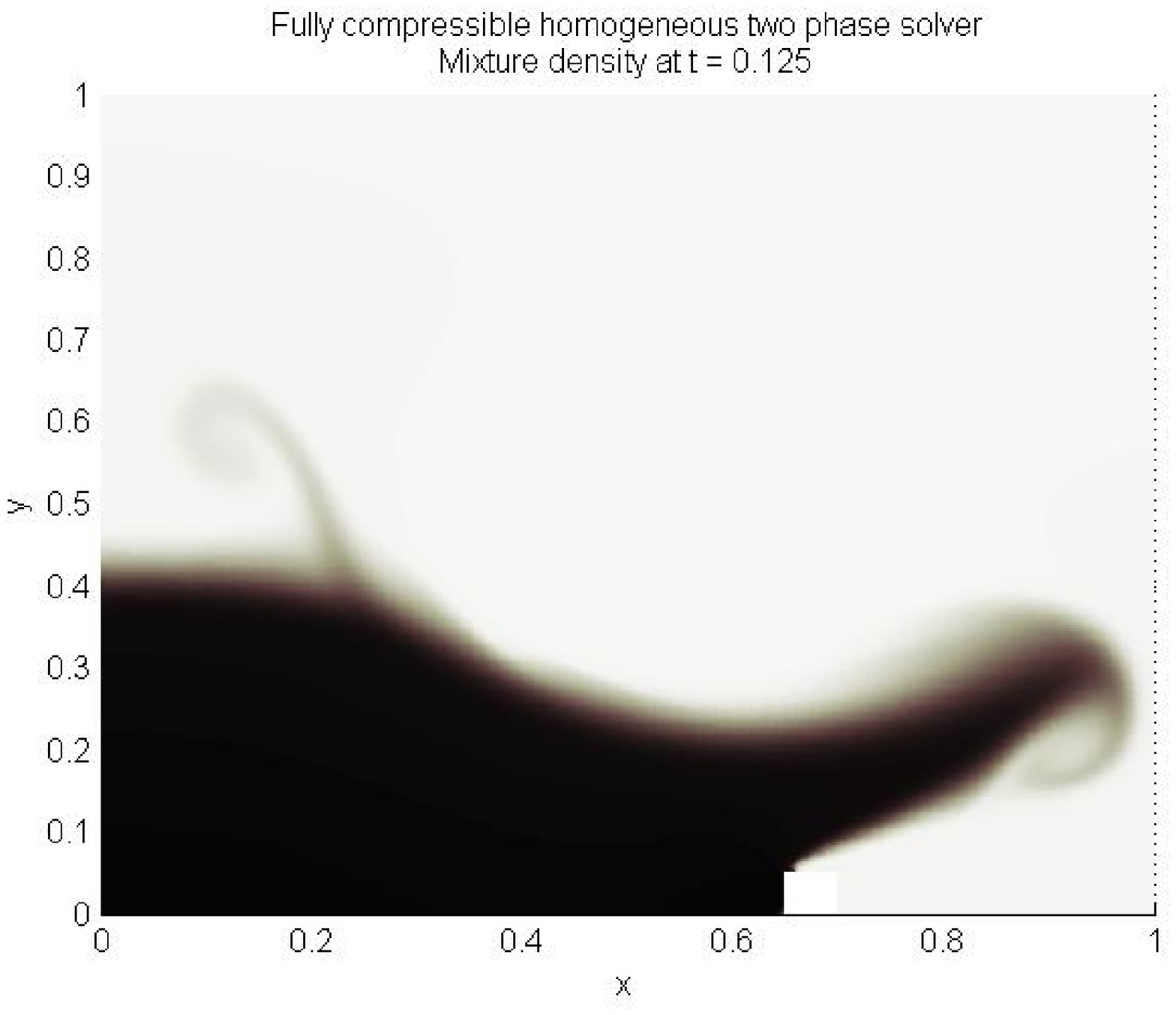}}
	\caption[Splash creation due to the interaction with step]{Splash creation due to the interaction with the step.}
\end{figure}


\begin{figure}
	\centering
	\subfigure[$t = 0.2$]%
	{\includegraphics[width=0.2\textwidth]{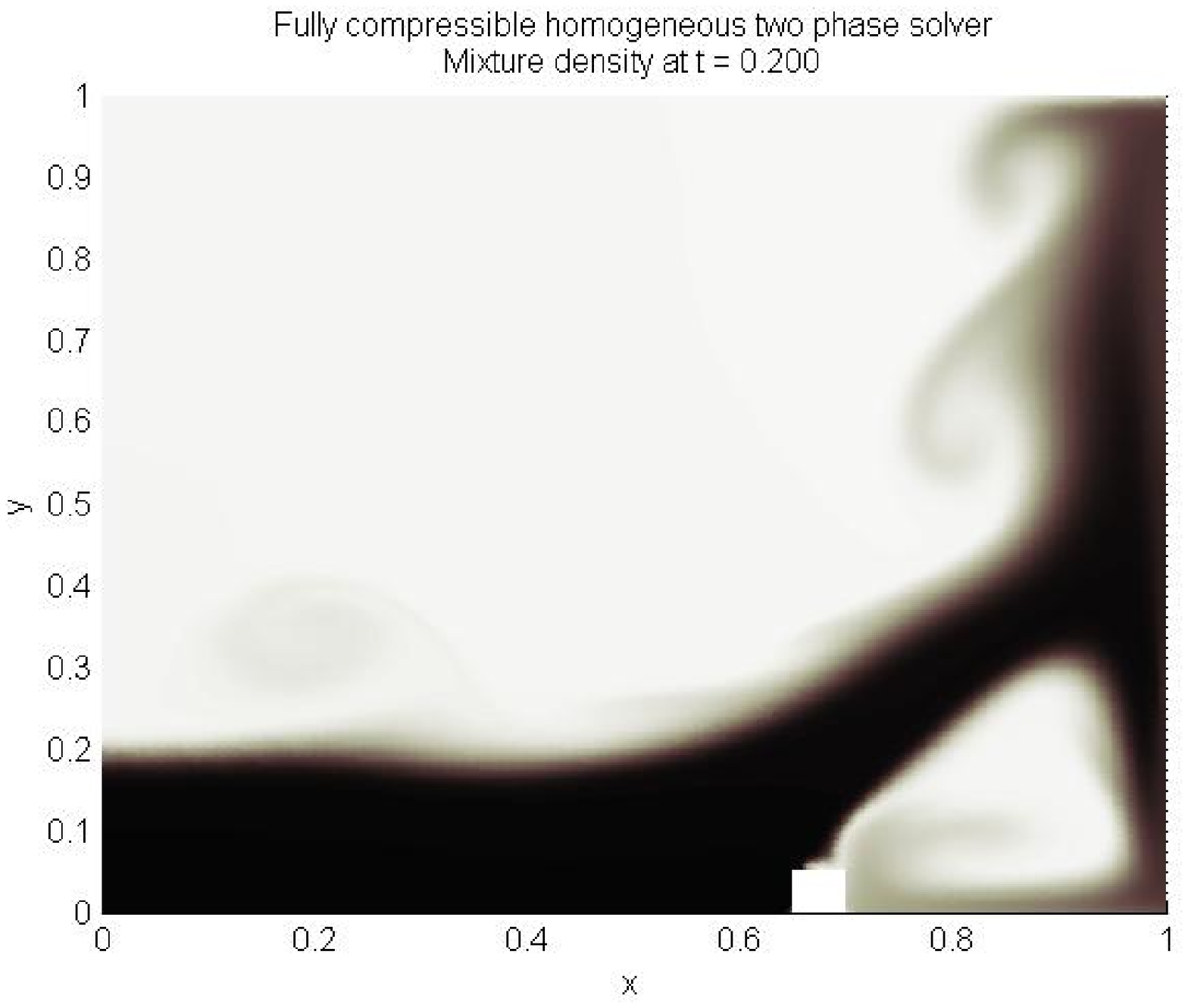}} \quad
	\subfigure[$t = 0.225$]%
	{\includegraphics[width=0.2\textwidth]{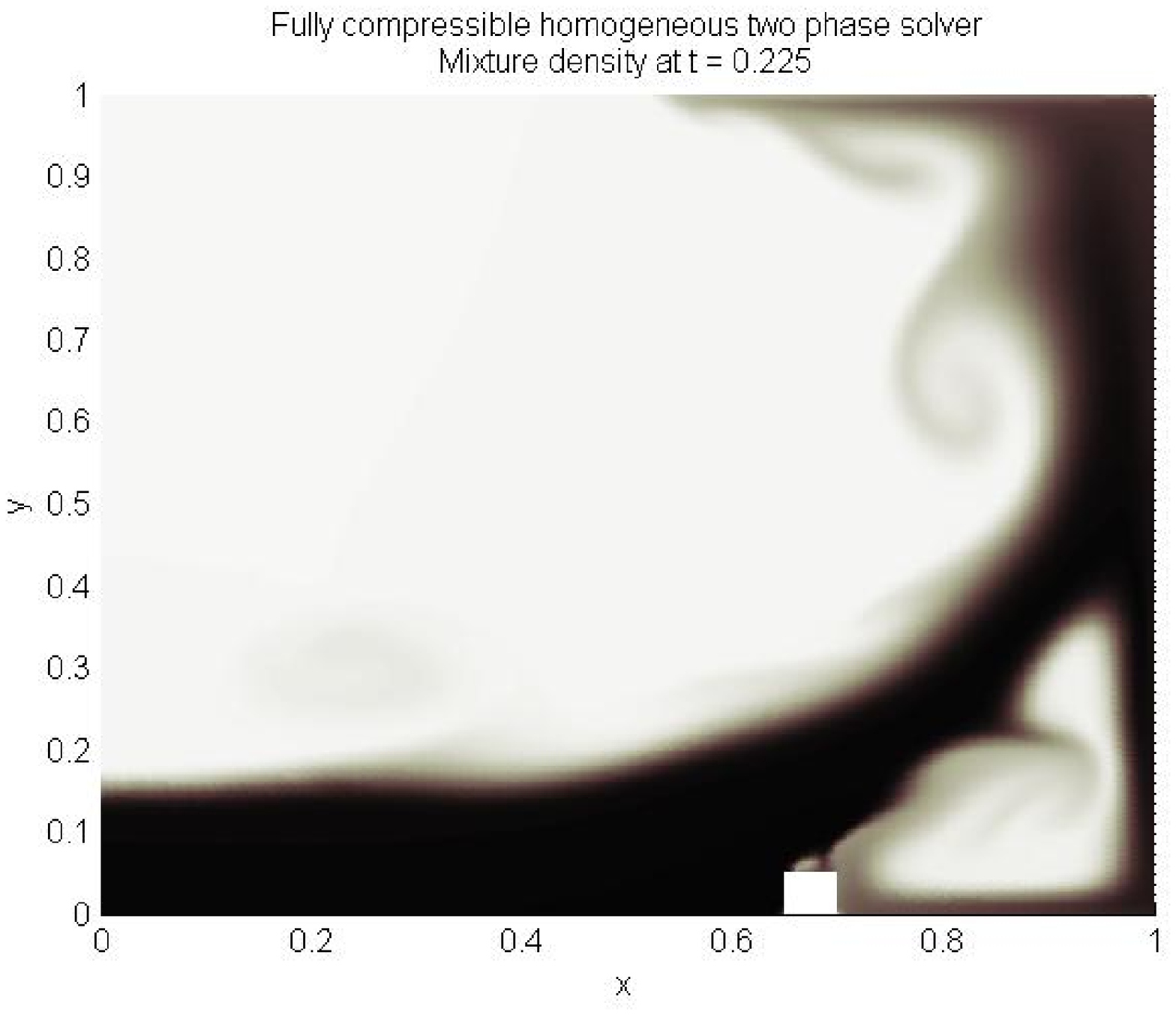}}
	\caption[Water strikes the wall - II]{The liquid hits the wall.}
\end{figure}

\begin{figure}
	\centering
	\subfigure[$t = 0.3$]%
	{\includegraphics[width=0.2\textwidth]{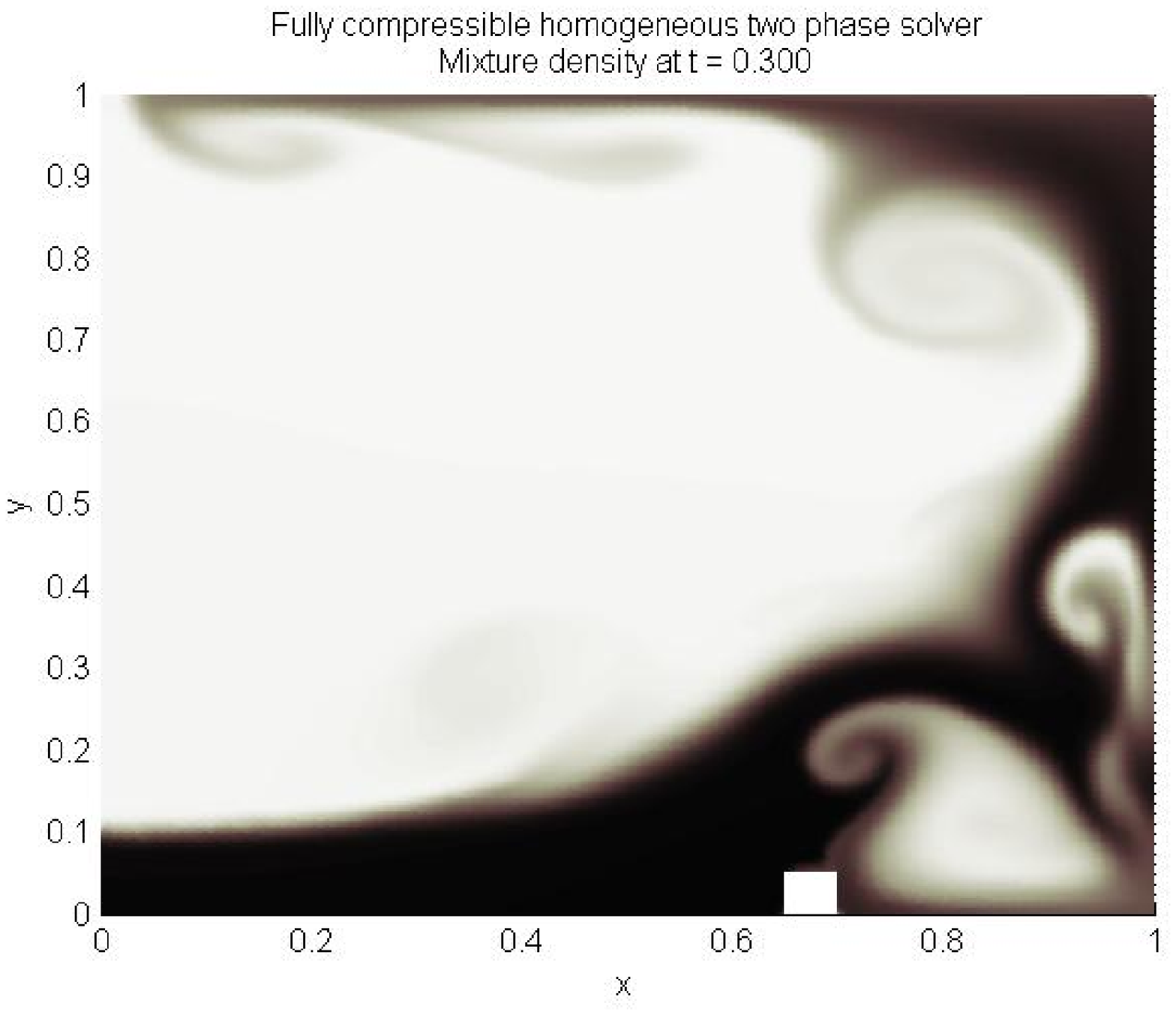}} \quad
	\subfigure[$t = 0.4$]%
	{\includegraphics[width=0.2\textwidth]{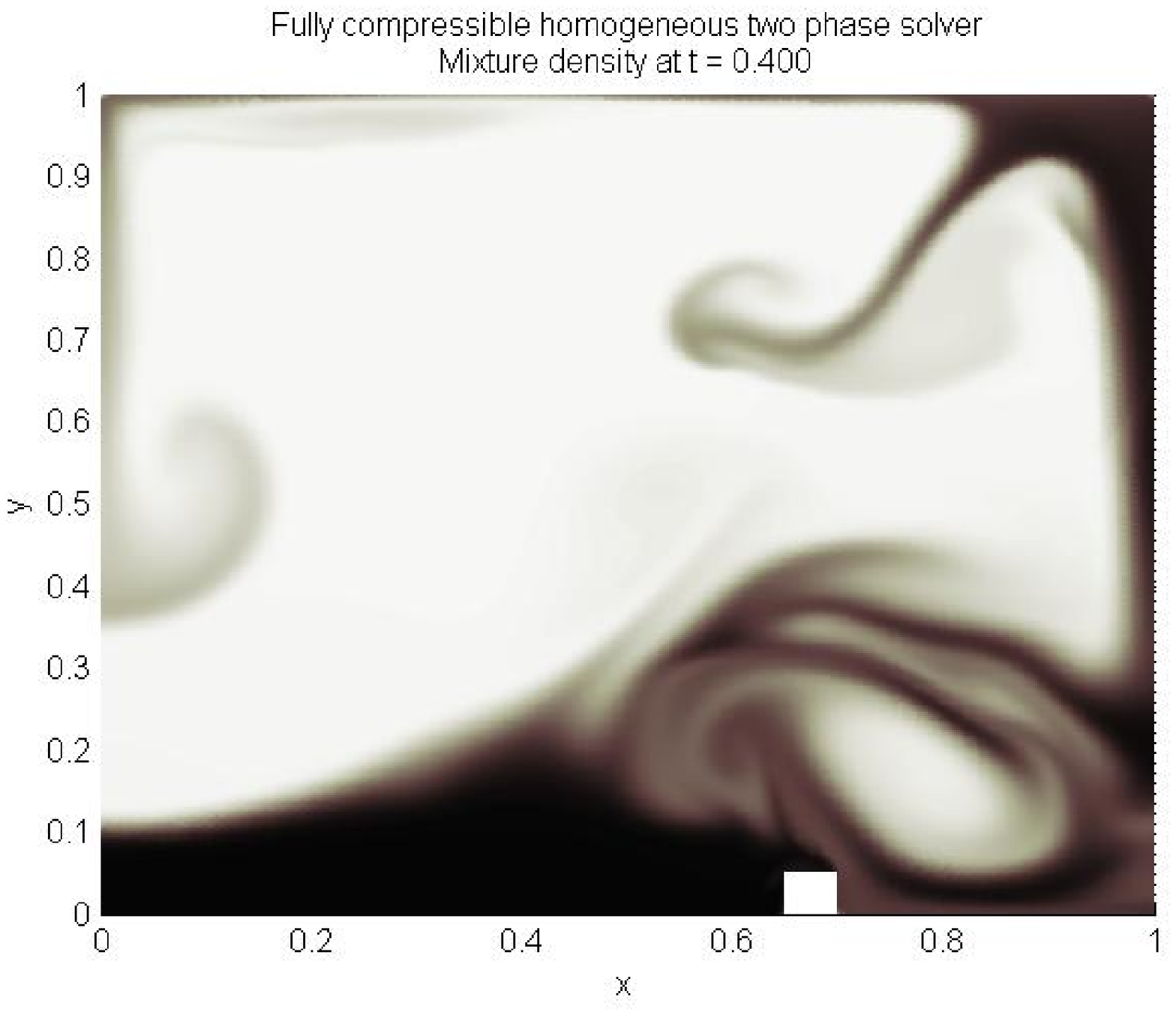}}
	\caption[Splash is climbing the wall]{The splash is climbing along the wall.}
\end{figure}

\section*{Conclusions}

In this article we presented a simple mathematical model for simulating water wave impacts. The preliminary results are encouraging and the validation of this approach will be the subject of future work. Namely, we are going to perform qualitative and quantitative comparisons with the more general six equations model \cite{Ishii1975}.

We also presented an efficient numerical approach for discretizing the governing equations. It is a second order finite volume scheme on unstructured meshes. This method was implemented in our research code. By construction, our code has excellent mass, momentum and energy conservation properties. Numerical tests presented in \cite{Dutykh2007a} partially validate the method.

We also plan to carry out a parametric study with our solver. The influence of aeration, gas properties and other factors on the impact pressures is very important for industrial applications.

\bibliographystyle{asmems4}
\bibliography{biblio}

\end{document}